\documentclass[aps,pra, twocolumn, groupedaddress, twoside]{revtex4-2}

\usepackage{graphicx}
\usepackage{amsfonts}
\usepackage{amsmath}
\usepackage{amssymb}

\usepackage{xcolor}
\definecolor{grey}{rgb}{0.7,0.7,0.7}
\definecolor{db}{rgb}{0,0,0.5}

\usepackage{fancyhdr}
 
\def\bs{\boldsymbol}
\def\bse{\begin{subequations}}
\def\ese{\end{subequations}}
\def\beqn{\begin{eqnarray}}
\def\eeqn{\end{eqnarray}}

\begin{document}



\title{\colorbox{db}{\parbox{\linewidth}{  \centering \parbox{0.9\linewidth}{ \textbf{\centering\Large{\color{white}{ \vskip0.2em{ Measurement of Orbital Angular Momentum of Light using Stokes Parameters and Barnett's Formalism}\vskip0.8em}}}}}}}



\author{Anirban Debnath}
\email[]{anirban.debnath090@gmail.com}
\affiliation{School of Physics, University of Hyderabad, Hyderabad 500046, India}

\author{Nirmal K. Viswanathan}
\email[]{nirmalsp@uohyd.ac.in}
\affiliation{School of Physics, University of Hyderabad, Hyderabad 500046, India}


\date{\today}

\begin{abstract}
\noindent{\color{grey}{\rule{0.784\textwidth}{1pt}}}
\vspace{-0.8em}

We present a formalism for experimental determination of the orbital angular momentum (OAM) of a paraxial optical beam-field by intertwining Barnett's formalism and Stokes parameter measurements. Using Barnett's formalism we calculate the OAM flux density of a suitably general 2-dimensional beam-field; and re-express it in terms of experimentally measurable quantities. Considering a demonstrative optical system, we obtain the OAM flux density profiles, total OAM flux and OAM per photon --- leading towards the experimental characterization of OAM using Barnett's formalism. Our method will find potential applications in nano-optical processes involving the transfer of electromagnetic OAM to particles and across interfaces.
{\color{grey}{\rule{0.784\textwidth}{1pt}}}
\end{abstract}


\maketitle



\tableofcontents

{\color{grey}{\noindent\rule{\linewidth}{1pt}}}


\section{Introduction} \label{Sec_Intro}

The presence of orbital angular momentum (OAM) in optical beams has first been identified three decades ago by Allen et al. \cite{AllenOAM1992, Berry1998, AllenPadgett2000}. In their work they have shown the decomposition of the total angular momentum (AM) of Laguerre-Gaussian (LG) laser modes to OAM and spin angular momentum (SAM). But subsequently, Barnett and Allen \cite{Barnett1994} have shown that such a decomposition is invalid for non-paraxial beams.

However, after almost a decade, Barnett \cite{Barnett2002} has identified that the AM density based formulations \cite{AllenOAM1992, Berry1998, AllenPadgett2000} are at the core of this invalidity. The AM density of an electromagnetic field is the cross product of the radius vector (from the axis of rotation) and the linear momentum density. This quantity does not give the true flow of AM across an interface. As explained by Barnett, the more physically meaningful quantity in this context is the AM flux density, which describes the true flow of AM. In his work, Barnett has established a complete AM flux density formalism, via which he has resolved the issues posed by AM density formulations. In particular, he has shown that the AM flux density can be consistently decomposed into OAM and SAM flux densities for all paraxial and non-paraxial cases.

For an LG mode, the mode order $l$ equals the topological charge $t$ \cite{NyeBerry1974, BH1977, OAMBook2013, Gbur}; and the OAM per photon is obtained as $l\hbar = t\hbar$. The topological charge can be obtained by simple diffraction or interference experiments \cite{Harris1994, Padgett1996, Soskin1997,  DovePrisms2002, SingleSlit, TriAperture2010, LaveryC13_OAM2013, Gbur}; and hence the OAM per photon can be straightforwardly determined. However, such a straightforward calculation cannot be performed for a non-canonical phase vortex \cite{Berry1998, MTerrizaChapterOAM2013}, which is generated by a superposition of multiple LG modes. A detailed OAM determination method is thus necessary for general vortex beam-fields.

Standard OAM measurement methods for general superposed vortices \cite{LitvinAziDecomp2012, SchulzeModalDecomp2012, SchulzeModalDecomp2013, Zhang2015} are formulated based on OAM density \cite{AllenOAM1992, Berry1998, AllenPadgett2000}. To our knowledge, an OAM measurement method based on its flux density is not present in the literature. In addition, the standard measurement methods for non-canonical vortices focus on the decomposition of the superposed states to the constituent eigenmodes \cite{LitvinAziDecomp2012, SchulzeModalDecomp2012, SchulzeModalDecomp2013}. While such a modal decomposition has its own significance in identifying and utilizing the constituent OAM eigenstates \cite{WangTerabit2012, ZhangEncoding2012, MairEntanglement2001, VaziriEntanglement2002}, it seems unnecessary to involve modal decomposition for the sole purpose of determining the total OAM transfer to particles and across interfaces.

In the present work we establish a formalism for experimental determination of the OAM of a paraxial optical beam-field by intertwining Barnett's AM flux density formalism and Stokes parameter measurements \cite{Goldstein}. 
We consider a general 2-dimensional beam-field, for which we calculate the OAM flux density via Barnett's formalism. Then integrating the flux density we obtain the total OAM flux across the beam cross-section. Finally, by comparing this with the total beam power, we obtain the required measure of the OAM per photon.

To demonstrate the measurement, we create an optical system --- both in the simulation and in the experiment --- based on a paraxial beam reflection at a plane isotropic dielectric interface. Simulated and experimental OAM flux density profiles are obtained and compared for this system.
Our computational analyses involve numerical differentiations and integrations of simulated and experimentally obtained profiles.
Agreements between the theoretical and experimental results are found; and to our knowledge, this is the first reporting of the OAM per photon measurement utilizing both Barnett's formalism and Stokes parameters together.
Our work thus represents a simplified yet elegant method of OAM measurement of general complex 2D beam-fields, which paves the way towards future applications involving OAM transfer to particles and across interfaces.

%


\section{Formalism} \label{Sec_Formalism}

\subsection{Theoretical Foundation}

A general 2D beam-field can be expressed as (suppressing the $kz - \omega t$ phase term)
\begin{equation} 
\boldsymbol{\mathcal{E}} = \boldsymbol{\mathcal{E}}_{x} + \boldsymbol{\mathcal{E}}_{y} = \mathcal{E}_{x} \, \hat{\mathbf{x}} + \mathcal{E}_{y} \, \hat{\mathbf{y}}; \label{E_def} 
\end{equation}
where, $\mathcal{E}_{x}$ and $\mathcal{E}_{y}$ are complex functions of $(x,y)$.
The OAM flux density of the field $\boldsymbol{\mathcal{E}}$ across the beam cross-section ($xy$ plane) is obtained by using Barnett's formalism \cite{Barnett2002, Gbur} as
\begin{equation}
M_{orb} = \dfrac{\epsilon_0}{2 k} \mathfrak{Im} \left[ (\partial_\phi \mathcal{E}_x) \mathcal{E}_x^* + (\partial_\phi \mathcal{E}_y) \mathcal{E}_y^* \right] ; \label{Morb_def}
\end{equation}
where, $\epsilon_0 = $ permittivity of empty space; $k = 2\pi/\lambda$ ($\lambda = $ free-space wavelength); $\partial_\phi = x \partial_y - y \partial_x$ is the partial differential operator with respect to the azimuthal coordinate variable $\phi$; and $\mathfrak{Im}(Z)$ represents the imaginary part of a complex function $Z$. Here we have considered the medium's refractive index $= 1$.

From Barnett's analysis, the above expression of $M_{orb}$ appears in the context of calculating the total OAM flux through the beam cross-section:
\begin{equation}
L_{orb} = \int_{-\infty}^\infty \int_{-\infty}^\infty M_{orb} \, dx \, dy. \label{Lorb_def}
\end{equation}
In this context, the $M_{orb}$ expression of Eq.~(\ref{Morb_def}) only gives the term which has a non-zero contribution to $L_{orb}$ for OAM carrying beams. 
Mathematically, it is possible to have an additional function $M(x,y)$ that satisfies
\begin{equation}
\int_{-\infty}^\infty \int_{-\infty}^\infty M \, dx \, dy = 0, \label{M_function}
\end{equation}
which can be added to the expression of Eq.~(\ref{Morb_def}) without affecting the $L_{orb}$ result of Eq.~(\ref{Lorb_def}).
This mathematical subtlety, however, does not affect the physical interpretation of $L_{orb}$ describing the total OAM transferred across the entire beam cross-section per unit time. So, for the present work, it is sufficient to proceed further by considering the 
$M_{orb}$ expression of Eq.~(\ref{Morb_def}).

The power $\mathcal{P}$ of the field $\boldsymbol{\mathcal{E}}$ is obtained by integrating its intensity $\mathcal{I} = (1/2 \mu_0 c) |\bs{\mathcal{E}}|^2$ ($\mu_0 =$ magnetic permeability of empty space; $c =$ speed of light in empty space):
\begin{equation}
\mathcal{P} = \int_{-\infty}^\infty \int_{-\infty}^\infty \mathcal{I} \, dx \, dy. \label{P_def}
\end{equation}
In a semiclassical description, this power is obtained due to $\mathcal{N} = \mathcal{P}/\hbar\omega$ photons (each of energy $\hbar\omega$) passing through the beam cross-section per unit time. The OAM per photon is then obtained as
\begin{equation}
l_{orb} = L_{orb}/\mathcal{N} = (L_{orb}/\mathcal{P}) \hbar\omega. \label{lorb_def}
\end{equation}

The above description is the theoretical foundation of our OAM determination method.
In order to use this in an experimental scenario, we first formulate a core scheme by considering a uniformly polarized field; and then extend it to a general complex 2D beam-field.

\subsection{The Measurement Scheme}

For the core scheme we consider a beam-field 
\begin{equation} 
\boldsymbol{\mathcal{E}} = \mathcal{E} e^{i\Phi} \, \hat{\mathbf{e}} \, ;
\hspace{1.5em}
\hat{\mathbf{e}} = a_{x} \, \hat{\mathbf{x}} + a_{y} e^{i\Phi_0} \, \hat{\mathbf{y}}, \hspace{1em} a_x^2 + a_y^2 = 1; \label{E1_def} 
\end{equation}
where, the amplitude $\mathcal{E}$ and phase $\Phi$ are real functions of $(x,y)$; whereas, $a_x$, $a_y$ and $\Phi_0$ are real quantities independent of $(x,y)$. Thus, $\mathcal{E} e^{i\Phi}$ denotes a spatially varying scalar wave term; and $\hat{\mathbf{e}}$ denotes an associated uniform polarization. 
It is to be noticed that, in any given situation, all phase terms are described with respect to an appropriately chosen fixed reference phase. Any spatially uniform phase term is cancelled in Eq.~(\ref{Morb_def}); and hence, the OAM flux calculation is not affected by the choice of the uniform reference phase.


The field $\boldsymbol{\mathcal{E}}$ of Eq.~(\ref{E1_def}) can be expressed in the general form of Eq.~(\ref{E_def}) by defining 
$\mathcal{E}_x = \mathcal{E} a_x e^{i\Phi}$ and $\mathcal{E}_y = \mathcal{E} a_y e^{i(\Phi + \Phi_0)}$. Using these in Eq.~(\ref{Morb_def}), we obtain the OAM flux density as
\begin{equation}
M_{orb} = \dfrac{\epsilon_0}{2 k} \mathcal{E}^2 (\partial_\phi \Phi) . \label{Morb1}
\end{equation}
Subsequently, by using Eqs.~(\ref{Lorb_def}), (\ref{P_def}), (\ref{lorb_def}) and (\ref{Morb1}), we obtain the final expression of the OAM per photon as
\begin{equation}
l_{orb} = \hbar \, \dfrac{\displaystyle\int_{-\infty}^\infty \int_{-\infty}^\infty \mathcal{I} \, (\partial_\phi \Phi) \, dx \, dy}{\displaystyle\int_{-\infty}^\infty \int_{-\infty}^\infty \mathcal{I} \, dx \, dy} . \label{lorb_final}
\end{equation}

The above equation summarizes the core scheme of our present formulation. This scheme is now to be extended to the case of a general inhomogeneously polarized complex beam-field 
\begin{equation}
\bs{\mathcal{E}}_{total} = \mathcal{E}_{0x} e^{i\Phi_x} \hat{\mathbf{x}} + \mathcal{E}_{0y} e^{i\Phi_y} \hat{\mathbf{y}}, 
\label{Etot}
\end{equation}
(real $\mathcal{E}_{0x}, \mathcal{E}_{0y}, \Phi_x, \Phi_y$). Considering $\mathcal{E}_{x} = \mathcal{E}_{0x} e^{i\Phi_x}$ and $\mathcal{E}_{y} = \mathcal{E}_{0y} e^{i\Phi_y}$ in Eq.~(\ref{Morb_def}), we get
\beqn
M_{orb} &=& \dfrac{\epsilon_0}{2 k} \left[ \mathcal{E}_{0x}^2 (\partial_\phi \Phi_x) + \mathcal{E}_{0y}^2 (\partial_\phi \Phi_y) \right] \nonumber \\
&=& M_{orb}^{(x)} + M_{orb}^{(y)} \hspace{1em} (\mbox{say}) . \label{Morb_tot}
\eeqn
The total OAM flux density is thus the sum of the OAM flux densities of the individual component fields $\mathcal{E}_{0x} e^{i\Phi_x} \hat{\mathbf{x}}$ and $\mathcal{E}_{0y} e^{i\Phi_y} \hat{\mathbf{y}}$. Defining $\mathcal{I}_u = (1/2 \mu_0 c) \mathcal{E}_{0u}^2$ ($u = x,y$), we obtain the $l_{orb}$ expressions [Eq.~(\ref{lorb_final})] of the component fields $\mathcal{E}_{0u} e^{i\Phi_u} \hat{\mathbf{u}}$ as
\begin{equation}
l_{orb}^{(u)} = \hbar \, \dfrac{\displaystyle\int_{-\infty}^\infty \int_{-\infty}^\infty \mathcal{I}_u \, (\partial_\phi \Phi_u) \, dx \, dy}{\displaystyle\int_{-\infty}^\infty \int_{-\infty}^\infty \mathcal{I}_u \, dx \, dy} . \label{lorb_final_comps}
\end{equation}
The $l_{orb}$ expression for the total field $\bs{\mathcal{E}}_{total}$ [Eq.~(\ref{Etot})] is then obtained as
\begin{eqnarray}
l_{orb}^{total} &=& \hbar \, \dfrac{\displaystyle\int_{-\infty}^\infty \int_{-\infty}^\infty \left[\mathcal{I}_x \, (\partial_\phi \Phi_x) + \mathcal{I}_y \, (\partial_\phi \Phi_y) \right] \, dx \, dy}{\displaystyle\int_{-\infty}^\infty \int_{-\infty}^\infty \left( \mathcal{I}_x + \mathcal{I}_y \right) \, dx \, dy} \nonumber\\ 
&=& \hbar \, \dfrac{\mathcal{P}_x \, l_{orb}^{(x)} + \mathcal{P}_y \, l_{orb}^{(y)}}{\mathcal{P}_x + \mathcal{P}_y} \, ; \label{lorb_final_total}
\end{eqnarray}
where, $\mathcal{P}_u$ are the powers corresponding to the intensities $\mathcal{I}_u$ [Eq.~(\ref{P_def})].

In our experiments, we obtain the $\mathcal{I}$ and $\Phi$ terms via intensity measurements and appropriately arranged Stokes measurements --- all performed by capturing beam profile data using a CCD camera. 
We then numerically calculate the $\partial_\phi \Phi$ terms; and then numerically integrate $\mathcal{I}$ and $\mathcal{I} \, (\partial_\phi \Phi)$ to determine $l_{orb}$ [Eqs.~(\ref{lorb_final}), (\ref{lorb_final_comps}), (\ref{lorb_final_total})]. 



\subsection{Computational Considerations}

The following technical aspects are considered in our computational data analysis.

\begin{enumerate}

\item The phase associated to a complex quantity $a + i b$ is given by $\Phi = \tan^{-1}(b/a)$. From a computational perspective, the inverse tangent function has a principal value range $[-\pi/2, \pi/2]$. 
So, in order to utilize a $2\pi$ phase range $(-\pi, \pi]$, we implement the following conditions for $a<0$ : (1) $-\pi < \Phi < -\pi/2$ for $b < 0$, (2) $\pi/2 < \Phi \leq 2\pi$ for $b \geq 0$. If a phase range larger than $2\pi$ is required, then we perform an appropriate phase unwrapping \cite{FFT1982}.

\item 
The computational data of a vortex phase profile $\Phi$, in addition to the actual vortex/vortices, contain $2N\pi$ ($N \in \mathbb{I}$) discontinuities corresponding to the pitch of the helical wavefront structure. Numerical calculation of $\partial_\phi \Phi$ at such discontinuities requires some specialized steps, which can be avoided by using the alternative expression
\begin{eqnarray} 
\partial_\phi \Phi &=& \partial_\phi [\tan^{-1} (\tan \Phi)] \nonumber \\
&=& \cos\Phi \, (\partial_\phi \sin\Phi) - \sin\Phi \, (\partial_\phi \cos\Phi) . \label{dPhi_SinCos}
\end{eqnarray}

\item The theoretical intensity $\mathcal{I} = (1/2 \mu_0 c) \mathcal{E}^2$ has the unit of W/m$^2$; whereas, the CCD camera captures the intensity data in terms of pixel values (e.g., 8-bit data). However, since $\mathcal{I}$ appears equivalently in both the numerators and the denominators of Eqs.~(\ref{lorb_final}), (\ref{lorb_final_comps}) and (\ref{lorb_final_total}), any factor of conversion between the pixel value unit and the W/m$^2$ unit would cancel. So the computational $\mathcal{I}$ data can be directly used in Eqs.~(\ref{lorb_final}), (\ref{lorb_final_comps}) and (\ref{lorb_final_total}) without any unit conversion.

\item All data obtained by the CCD camera are 2D arrays with a Cartesian coordinate system structure. So we necessarily expand the operator $\partial_\phi$ in the form $\partial_\phi = x \partial_y - y \partial_x$ while computing the numerical derivatives.

\end{enumerate}


\section{Stokes Parameter Measurements} \label{Sec_Stokes}

\subsection{The Context}

Experimental techniques based on aperture and slit diffraction \cite{SingleSlit, TriAperture2010} are well-known to give the topological charges \cite{NyeBerry1974, BH1977, OAMBook2013, Gbur} of phase vortices. However, for a non-canonical vortex \cite{Berry1998, MTerrizaChapterOAM2013}, which is obtained via a superposition of different vortex eigenmodes, the topological charge information is not sufficient to reproduce the complete phase profile. Because of this, the aperture and slit diffraction methods are not sufficient to give the actual measure of the OAM of a general field function which is not an OAM eigenmode. We need the complete phase profile $\Phi$ to be used in Eq.~(\ref{lorb_final}) 
in order to determine the OAM.

A widely used method for identifying the variable phase structure of a beam-field is to superpose it with a reference plane-wave beam and to observe the interference pattern \cite{Harris1994, Padgett1996, Soskin1997, FFT1982, Zhao2017}. However, counting the number of forks in a forked interference pattern simply gives the topological charge, which is not sufficient for our purpose, as established already.
A more elaborate methodology is to reproduce the phase profile from the interference fringe patterns \cite{FFT1982, Zhao2017}; but it takes significant additional computational and/or experimental efforts. 

On this premise we rely on Stokes parameters \cite{Goldstein} to obtain the complete phase information. Of course, an overall phase profile $\Phi$, as in Eq.~(\ref{E1_def}), cannot be obtained by direct Stokes parameter measurements on the field $\bs{\mathcal{E}}$. So we summarize below a few ways to achieve the goal via appropriately engineering the measurable field.

\subsection{Phase Measurements} \label{Subsec_Stokes_PhaseMeasure}

For most cases we still use a reference plane-wave beam for our purpose, but we polarize it orthogonally as compared to the original field.

\begin{enumerate}

\item If $\hat{\mathbf{e}}$ [Eq.~(\ref{E1_def})] is a uniform linear polarization ($\Phi_0 = 0, \pi$), then we choose a coordinate system where $\hat{\mathbf{e}} = \hat{\mathbf{y}}$; and then superpose an $\hat{\mathbf{x}}$-polarized plane-wave beam with the original field $\bs{\mathcal{E}}$. The phase $\Phi$ is then given by the phase $\Phi_{23}$ of the complex Stokes parameter $S_2 + i S_3$.


\item If $\hat{\mathbf{e}}$ is a uniform circular or elliptical polarization, then we transform it to a linear polarization by using an appropriately oriented quarter wave plate (QWP); and then use the above method 1 to extract its phase information.

\item For a general inhomogeneously polarized complex beam-field $\bs{\mathcal{E}}_{total}$ [Eq.~(\ref{Etot})], we need the $\Phi_x$ and $\Phi_y$ phase profiles individually, to be used in Eqs.~(\ref{Morb_tot})--(\ref{lorb_final_total}). 
So we isolate the component fields $\mathcal{E}_{0x} e^{i\Phi_x} \hat{\mathbf{x}}$ and $\mathcal{E}_{0y} e^{i\Phi_y} \hat{\mathbf{y}}$ by using appropriate orientations of a polarizer. The phase structures of each of these component fields are then determined by using the method 1 above.



\end{enumerate}

While the above methods rely on plane wave superpositions, it is also possible to utilize known special properties and symmetries of the field under consideration such that its phase characteristics can be interpreted without requiring a reference plane wave. An example of this process is given in Sec.~\ref{Sec_Demo}.


\section{Demonstration} \label{Sec_Demo}

We choose a rather unconventional optical system to demonstrate our OAM measurement method. The significance of this approach is not only to demonstrate the measurement itself, but also (1) to unravel and establish the understanding that 
any complex beam-field, irrespective of its method of generation, can potentially possess explicitly quantifiable and experimentally measurable OAM; and (2) to demonstrate a case where a known property of the field is utilized to extract the phase information, without requiring a reference plane wave, as mentioned in Sec.~\ref{Subsec_Stokes_PhaseMeasure}.

\subsection{The Optical System} \label{Subsec_System}

\begin{figure*}
\begin{center}
\includegraphics[width = \linewidth]{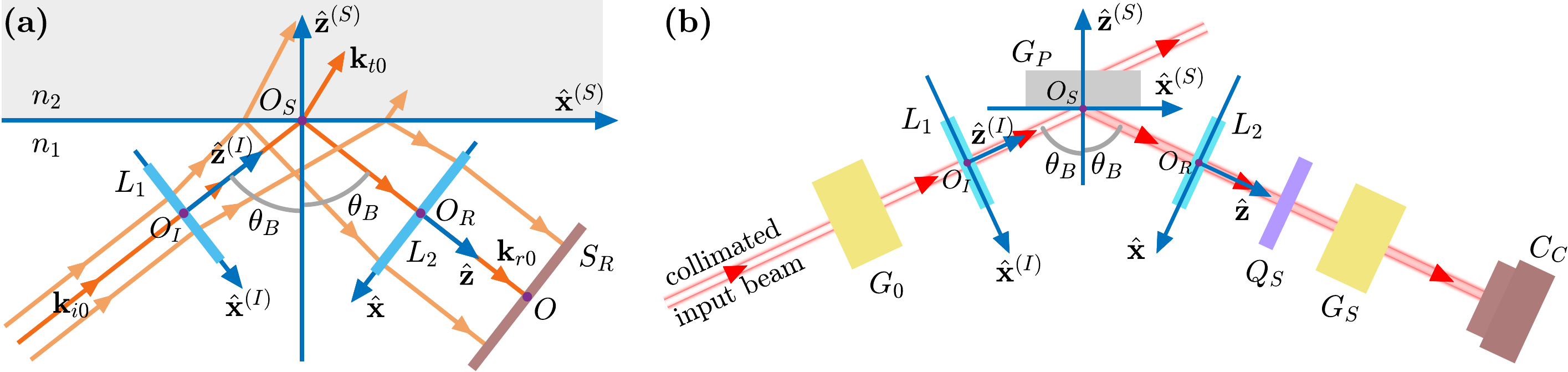}
\end{center}
\caption{The \textbf{(a)} simulated system and \textbf{(b)} experimental setup. The medium of incidence/reflection has a refractive index $n_1$. The glass plate $G_P$ has a refractive index $n_2$. The optical components are: $L_1 = $ diverging lens; $L_2 = $ collimating lens; $G_0, G_S = $ Glan-Thompson polarizers (GTP); $Q_S = $ quarter wave plate (QWP). A CCD camera $C_C$ represents the screen $S_R$. The coordinate systems $I(x^{(I)},y^{(I)},z^{(I)})$, $R(x,y,z)$ and $S(x^{(S)},y^{(S)},z^{(S)})$ are defined in reference to the central incident wavevector $\mathbf{k}_{i0}$, central reflected wavevector $\mathbf{k}_{r0}$, and the surface normal ($-\hat{\mathbf{z}}^{(S)}$).
} 
{\color{grey}{\rule{\linewidth}{1pt}}}
\label{Fig_Sys}
\end{figure*}

We consider a simulated optical system as shown in Fig.~\ref{Fig_Sys}(a), where an initial collimated Gaussian beam with a uniform linear polarization is diverged through a lens $L_1$, and subsequently reflected at a plane isotropic dielectric interface ($z^{(S)} = 0$), with the central angle of incidence being the Brewster angle $\theta_B$. The reflected diverging beam is collimated by using a lens $L_2$, and is subsequently observed at a screen $S_R$. 

The experimental realization of the system is shown in Fig.~\ref{Fig_Sys}(b); where, a Glan-Thompson polarizer (GTP) $G_0$ is used to set the initial uniform linear polarization of the collimated beam; the surface of a glass plate $G_P$ is used as the plane dielectric interface; and a CCD camera $C_C$ is used as the screen $S_R$. A QWP $Q_S$ and a GTP $G_S$ are used before $C_C$ to perform Stokes parameter measurements.


For the purpose of the present paper, we set the initial uniform linear polarization direction along $\hat{\mathbf{x}}^{(I)}$ [Fig.~\ref{Fig_Sys}(a)], which is the transverse magnetic (TM) polarization. However, the field profile distorts after passing through the diverging lens $L_1$ 
in such a way that only the central wavevector is incident at $\theta_B$ with TM polarization. This condition is not satisfied by the surrounding field; and hence a complicated reflection occurs.
The reflected field thus obtained is still locally linearly polarized at each point, but the polarization direction globally varies over the beam cross-section. This induced polarization inhomogeneity is preserved even after the beam is collimated by the lens $L_2$. 


We have simulated the above process \cite{ADNKVrt2020, ADNKVBrew2021}, and have obtained a final collimated field profile ($\mathcal{E}_{Rx}$ and $\mathcal{E}_{Ry}$ expressions are approximated via 2D surface fitting in MATLAB)
\bse \label{ER_Brew}
\begin{eqnarray}
& \boldsymbol{\mathcal{E}}_R = \mathcal{E}_{Rx} \, \hat{\mathbf{x}} + \mathcal{E}_{Ry} \, \hat{\mathbf{y}}; & \label{ERdef} \\
& \mathcal{E}_{Rx} = A_1 (x/w_R) G_R , \hspace{1em} \mathcal{E}_{Ry} = A_2 (y/w_R) G_R; & \label{ERxy}
\end{eqnarray}
\ese
where, $A_1$ and $A_2$ are field amplitude terms ($A_1, A_2 > 0$), 
$w_R$ is the effective half-width of the output collimated beam, and $G_R = e^{-\rho^2/w_R^2}$ ($\rho = \sqrt{x^2 + y^2}$) is an overall Gaussian envelope. The field functions $\mathcal{E}_{Rx}$ and $\mathcal{E}_{Ry}$ are thus first order Hermite-Gaussian (HG) modes. 

\begin{figure*}
\begin{center}
\includegraphics[width = 0.9\linewidth]{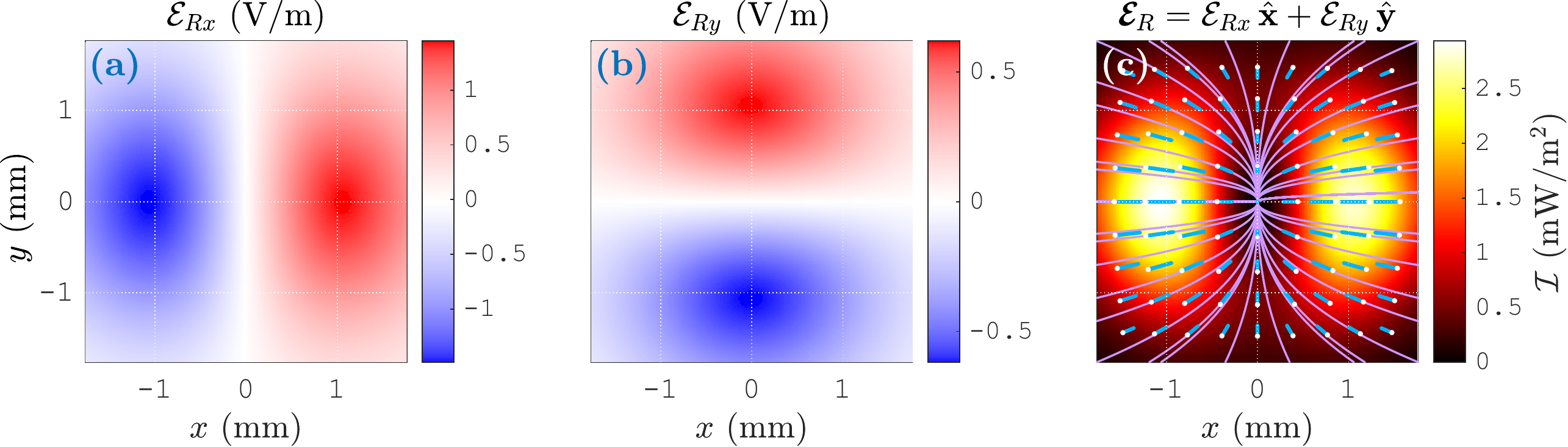}
\end{center}
\caption{Simulated profiles $\mathcal{E}_{Rx}$, $\mathcal{E}_{Ry}$ and $\boldsymbol{\mathcal{E}}_R$ [Eqs.~(\ref{ER_Brew})] for the chosen parameters. These profiles are not generated by using Eq.~(\ref{ERxy}); rather, these are obtained as exact simulation results, 
and the approximate forms of Eq.~(\ref{ERxy}) are subsequently obtained
via 2D surface fitting. In (c), the light blue line segments represent local linear polarizations, the white dots represent the `tips' of the field vectors at time $t = 0$, and the purple streamlines represent the orientation patterns of the linear polarizations.
} 
{\color{grey}{\rule{\linewidth}{1pt}}}
\label{Fig_Eprofiles}
\end{figure*}

We choose the following parameter values for the simulation and the experiment: laser power $1$ mW, free-space wavelength $\lambda = 632.8$ nm; refractive indices $n_1 = 1$, $n_2 = 1.52$; input beam half-width $w_0 = 0.6$ mm; focal length of lens $L_1$ $ = -5$ cm, focal length of lens $L_2$ $ = 12.5$ cm; propagation path-lengths $O_I O_S = 5$ cm, $O_S O_R = 2.5$ cm. The simulated $\mathcal{E}_{Rx}$, $\mathcal{E}_{Ry}$ and $\boldsymbol{\mathcal{E}}_R$ profiles are shown in Fig.~\ref{Fig_Eprofiles}. The $A_1$ and $A_2$ amplitude terms [Eq.~(\ref{ERxy})] for the simulated profiles are computationally obtained as $A_1 \approx 3.42239$ V/m and $A_2 \approx 1.44292$ V/m, whereas $w_R$ is obtained as $1.5$ mm.

\subsection{Phase Singularity Characteristics} \label{Subsec_Vortex}


\begin{figure*}
\begin{center}
\includegraphics[width = \linewidth]{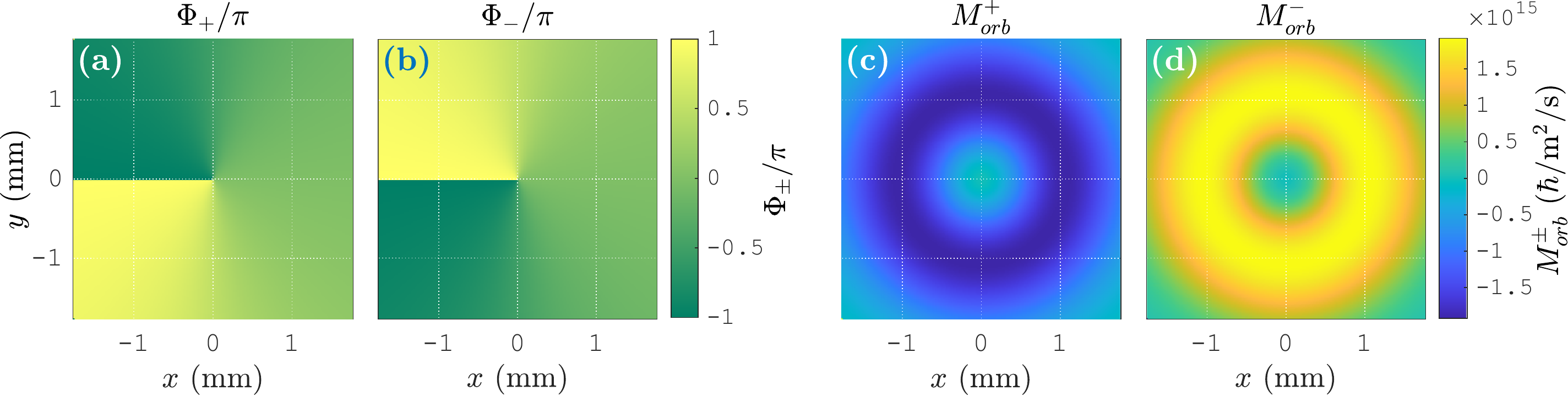}
\end{center}
\caption{Simulated \textbf{(a), (b)} phase profiles $\Phi_\pm$ [Eq.~(\ref{PhiPM_Brew})] and \textbf{(c), (d)} OAM flux density profiles $M_{orb}^\pm$ [Eq.~(\ref{MorbPM_Brew})] of the fields $\bs{\mathcal{E}}_{R\pm}$ [Eqs.~(\ref{Epm_BrewFull})].
} 
{\color{grey}{\rule{\linewidth}{1pt}}}
\label{Fig_SimPhiMorb}
\end{figure*}

To understand the inherent OAM characteristics of the field $\boldsymbol{\mathcal{E}}_R$, we re-express $\boldsymbol{\mathcal{E}}_R$ in terms of the $\hat{\boldsymbol{\sigma}}^\pm = (\hat{\mathbf{x}} \pm i \hat{\mathbf{y}})/\sqrt{2}$ spin-polarized component fields as
\bse \label{Epm_BrewFull}
\begin{eqnarray}
& \boldsymbol{\mathcal{E}}_R = \boldsymbol{\mathcal{E}}_{R+} + \boldsymbol{\mathcal{E}}_{R-} = \mathcal{E}_{R+} \, \hat{\boldsymbol{\sigma}}^+ + \mathcal{E}_{R-} \, \hat{\boldsymbol{\sigma}}^- ; & \label{E=Ep+Em} \\
& \mathcal{E}_{R\pm} = \left( G_R / \sqrt{2} w_R \right) (A_1 x \mp i A_2 y) = \mathcal{E}_{R0} \, e^{i\Phi_{\pm}}; \label{Epm_BrewExp} & \\
& \mathcal{E}_{R0} = \left( G_R / \sqrt{2} w_R \right) \left( A_1^2 x^2 + A_2^2 y^2 \right)^\frac{1}{2}; \label{ER0_Brew} & \\
& \Phi_\pm = \mp \tan^{-1} [(A_2/A_1) \tan\phi]; \label{PhiPM_Brew} &
\end{eqnarray}
\ese
where, $\mathcal{E}_{R0}$ and $\Phi_\pm$ are respectively the field amplitude and phase functions.
The simulated $\Phi_\pm$ profiles, for the presently considered parameters, are shown in Figs.~\ref{Fig_SimPhiMorb}(a) and \ref{Fig_SimPhiMorb}(b).
Clearly, at the origin, $\mathcal{E}_{R0} = 0$ and $\Phi_\pm$ are singular. The fields $\boldsymbol{\mathcal{E}}_{R\pm}$ are thus vortex beam-fields carrying OAM. In the present context these $\boldsymbol{\mathcal{E}}_{R\pm}$ fields serve as the $\boldsymbol{\mathcal{E}}$ fields of Eq.~(\ref{E1_def}) ($\mathcal{E}_{R0} \equiv \mathcal{E}$, $\Phi_\pm \equiv \Phi$, $\hat{\boldsymbol{\sigma}}^\pm \equiv \hat{\mathbf{e}}$), whose OAM per photon we intend to determine.

The topological charges of the $\Phi_\pm$ phase singularities, which are defined in the form \cite{NyeBerry1974, BH1977, OAMBook2013, Gbur}
\begin{equation}
t = \dfrac{1}{2\pi} \oint_\mathcal{C} \boldsymbol{\nabla} \Phi(\mathbf{r}) \cdot d\mathbf{r}, 
\end{equation}
($\mathcal{C} =$ closed contour around the singularity), are obtained here as $t_\pm = \mp 1$ --- physically signifying that, if $\phi$ is increased by $2\pi$, then the phases $\Phi_\pm$ change by $\mp 2\pi$.
However, here $\Phi_\pm$ are not proportional to $\phi$ (since $A_1 \neq A_2$), and hence the associated phase vortices are non-canonical \cite{MTerrizaChapterOAM2013}; i.e. the field functions $\mathcal{E}_{R\pm}$ are not pure LG modes. 
If they were pure LG modes, 
then one would have identified their mode orders as $l_\pm = t_\pm = \mp 1$; which would straightforwardly give their OAM per photon as $\mp\hbar$. This straightforwardness is, however, not applicable to non-canonical vortices, and hence an explicit determination of their OAM is required. 
In the following subsections we carry this out by using our method established in Sec.~\ref{Sec_Formalism}, both analytically and experimentally; and also verify the results by using a modal decomposition.

\subsection{OAM Determination using Eq.~(\ref{lorb_def})} \label{Subsec_OAMcalcEq}

Using Eqs.~(\ref{ER0_Brew}) and (\ref{PhiPM_Brew}) in Eq.~(\ref{Morb1}), we obtain the OAM flux densities of the $\boldsymbol{\mathcal{E}}_{R\pm}$ component fields as
\begin{equation}
M_{orb}^\pm = \mp \dfrac{\epsilon_0}{2 k} \dfrac{A_1 A_2}{2} \dfrac{\rho^2}{w_R^2} G_R^2. \label{MorbPM_Brew}
\end{equation}
Simulated $M_{orb}^\pm$ profiles for the presently considered system parameters are shown in Figs.~\ref{Fig_SimPhiMorb}(c) and \ref{Fig_SimPhiMorb}(d). 
Using Eq.~(\ref{MorbPM_Brew}) in Eq.~(\ref{Lorb_def}), we obtain the OAM fluxes 
\begin{equation}
L_{orb}^\pm = \mp \dfrac{\epsilon_0}{2 k} A_1 A_2 \dfrac{\pi w_R^2}{8}. \label{LorbPM_Brew}
\end{equation}
The powers of the $\boldsymbol{\mathcal{E}}_{R\pm}$ fields are obtained by using Eqs.~(\ref{P_def}) as
\begin{equation}
\mathcal{P}_\pm = \dfrac{1}{2 \mu_0 c} \displaystyle\int_{-\infty}^\infty \! \displaystyle\int_{-\infty}^\infty \!\! |\mathcal{E}_{R\pm}|^2 \, dx \, dy = \dfrac{1}{2 \mu_0 c} \dfrac{A_1^2 + A_2^2}{2} \dfrac{\pi w_R^2}{8}. 
\end{equation}
Hence, the OAM per photon corresponding to $\boldsymbol{\mathcal{E}}_{R\pm}$ are obtained by using Eq.~(\ref{lorb_def}) as
\begin{equation}
l_{orb}^\pm = \dfrac{L_{orb}^\pm}{\mathcal{P}_\pm}  \hbar \omega = \mp \dfrac{2 A_1 A_2}{A_1^2 + A_2^2} \hbar \, . \label{lorbpm_Brew}
\end{equation}
Thus, in the total beam-field $\boldsymbol{\mathcal{E}}_R$ [Eqs.~(\ref{ER_Brew}), (\ref{Epm_BrewFull})], a photon with spin $\hat{\boldsymbol{\sigma}}^\pm$ carries an OAM $l_{orb}^\pm$ given by Eq.~(\ref{lorbpm_Brew}).
Since $A_1 \neq A_2$, these $l_{orb}^\pm$ OAM are fractions of $\mp \hbar$, which is a remarkable result. 
These fractional OAM are unrelated to any fractional vortex \cite{Vasnetsov1998, BerryFracVortex2004}, because the topological charges of the presently considered vortices are $t_\pm = \mp 1$. Instead, these are related to the non-canonical nature \cite{Berry1998, MTerrizaChapterOAM2013} of the vortices of the phases $\Phi_\pm$ [Eq.~(\ref{PhiPM_Brew})].

\subsection{Verification using Modal Decomposition} \label{Subsec_OAMcalcModal}

Equation~(\ref{E=Ep+Em}) represents a non-separable state in a semiclassical single-photon picture, which implies that, if a single photon is detected in a polarization state $\hat{\boldsymbol{\sigma}}^\pm$, it is also readily identified to be in a spatial state $\mathcal{E}_{R\pm} = \mathcal{E}_{R0} \, e^{i\Phi_{\pm}}$. 
Using $x = \rho \cos\phi$ and $y = \rho \sin\phi$ in Eq.~(\ref{Epm_BrewExp}), and rearranging the terms, we obtain the modal decompositions of the states $\mathcal{E}_{R\pm}$ as
\begin{subequations} \label{Epm=B12_superpos}
\begin{eqnarray}
& \mathcal{E}_{R\pm} = \left( \rho \, G_R / \sqrt{2} w_R \right) \left( B_1 e^{\mp i\phi} + B_2 e^{\pm i\phi} \right); & \\
& B_1 = (A_1 + A_2)/2, \hspace{1em} B_2 = (A_1 - A_2)/2. &
\end{eqnarray}
\end{subequations}
The states $\mathcal{E}_{R\pm}$ are thus linear combinations of pure LG modes with $l = \pm 1$, with coefficient amplitude terms $B_1$ and $B_2$. The LG modes with $l = \pm 1$ have OAM per photon $\pm \hbar$. The expectation values of the OAM per photon associated to the $\mathcal{E}_{R\pm}$ states are then obtained as
\begin{equation}
l_{orb}^\pm = \dfrac{B_1^2 (\mp \hbar) + B_2^2 (\pm \hbar)}{B_1^2 + B_2^2} = \mp \dfrac{2 A_1 A_2}{A_1^2 + A_2^2} \hbar \, ; \label{lorbpm_Brew_Photon}
\end{equation}
which is the same result as in Eq.~(\ref{lorbpm_Brew}). In this way the fractional OAM $l_{orb}^\pm$ is explained and verified via modal decomposition.

The decomposition in Eqs.~(\ref{Epm=B12_superpos}) seems quite straightforward, and it raises the question why our proposed approach of Sec.~\ref{Subsec_OAMcalcEq} is necessary. 
However, Eqs.~(\ref{Epm=B12_superpos}) owe their simplicity to the presently considered simple field functions $\mathcal{E}_{R\pm}$ [Eq.~(\ref{Epm_BrewExp})].
For a general beam-field, such a modal decomposition is extremely complicated both analytically and experimentally \cite{LitvinAziDecomp2012, SchulzeModalDecomp2012, SchulzeModalDecomp2013}. Modal decomposition has its own significant purposes in the utilization of the constituent OAM eigenstates in novel applications such as classical communication and entanglement of OAM states \cite{WangTerabit2012, ZhangEncoding2012, MairEntanglement2001, VaziriEntanglement2002}. However, the sole purpose of determining the OAM, in applications involving OAM transfer to particles or across interfaces, can be served by using our formalism without ever requiring to perform a modal decomposition --- which is a significant advantage offered by our formalism. 

\subsection{Experimental OAM Measurement} \label{Subsec_OAMexp}

\begin{figure*}
\begin{center}
\includegraphics[width = \linewidth]{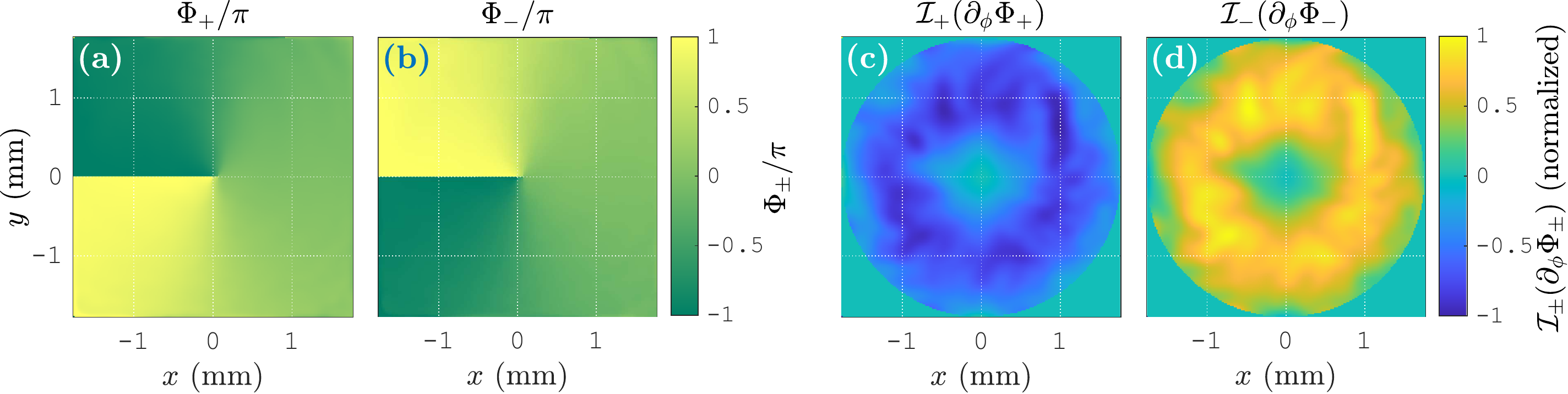}
\end{center}
\caption{\textbf{(a), (b)} Experimental $\Phi_\pm$ profiles, corresponding to the simulated profiles of Figs.~\ref{Fig_SimPhiMorb}(a), \ref{Fig_SimPhiMorb}(b). The smoothness of these data are achieved via computationally reducing their noise; and these data are subsequently used in computing $\partial_\phi \Phi_\pm$.
\textbf{(c), (d)} Experimental $\mathcal{I}_\pm (\partial_\phi \Phi_\pm)$ profiles, corresponding to the simulated $M_{orb}^\pm$ profiles of Figs.~\ref{Fig_SimPhiMorb}(c), \ref{Fig_SimPhiMorb}(d) (via Eq.~(\ref{Morb1}), $\mathcal{I}_\pm (\partial_\phi \Phi_\pm) \propto M_{orb}^\pm$). 
It is to be noticed that the computational noise reduction makes the experimental $\Phi_\pm$ profiles smooth enough so that the remaining noise is not visually understandable; but the subsequent numerical differentiation enhances the noise significantly.
} 
{\color{grey}{\rule{\linewidth}{1pt}}}
\label{Fig_ExpPhiMorb}
\end{figure*}

The experimental measures of $l_{orb}^\pm$ are obtained from Eq.~(\ref{lorb_final}) as 
\begin{equation}
l_{orb}^\pm = \hbar \, \dfrac{\displaystyle\int_{-\infty}^\infty \int_{-\infty}^\infty \mathcal{I}_\pm \, (\partial_\phi \Phi_\pm) \, dx \, dy}{\displaystyle\int_{-\infty}^\infty \int_{-\infty}^\infty \mathcal{I}_\pm \, dx \, dy} , \label{lorbPM_Exp}
\end{equation}
where, $\mathcal{I}_\pm = (1/2\mu_0 c) \, |\bs{\mathcal{E}}_{R\pm}|^2 = (1/2\mu_0 c) \, \mathcal{E}_{R0}^2$ [Eqs.~(\ref{Epm_BrewFull})]. To obtain the $\mathcal{I}_\pm$ profiles experimentally, the $\bs{\mathcal{E}}_{R\pm}$ component fields are individually extracted by orienting the fast axis of $Q_S$ [Fig.~\ref{Fig_Sys}(b)] along $\hat{\mathbf{y}}$, and then orienting the transmission axis of $G_S$ along $\hat{\mathbf{d}}^\pm = (\hat{\mathbf{x}} \pm \hat{\mathbf{y}})/\sqrt{2}$. 
By arranging additional appropriate orientations of $Q_S$ and $G_S$, and by analyzing the resulting intensity profiles, the Stokes parameter profiles \cite{Goldstein} are obtained.
The phase difference $\Delta\Phi = \Phi_- - \Phi_+$ is then given by the phase $\Phi_{12}$ of the complex Stokes parameter $S_1 + i S_2$ \cite{BliokhRev2019}. At this point we use a known special property of the considered system --- $\Phi_+ + \Phi_- = 0$ [Eq.~(\ref{PhiPM_Brew})] --- to obtain the individual phases as $\Phi_\pm = \mp\Delta\Phi/2$. The experimentally obtained $\Phi_\pm$ profiles are shown in Figs.~\ref{Fig_ExpPhiMorb}(a) and \ref{Fig_ExpPhiMorb}(b), which match well with the corresponding simulated profiles of Figs.~\ref{Fig_SimPhiMorb}(a) and \ref{Fig_SimPhiMorb}(b) respectively. This process thus shows an example case where the phase information is extracted by utilizing a special system property, along with using Stokes parameter measurements, but without using any reference plane wave superposition.

The corresponding experimentally obtained $\mathcal{I}_\pm \, (\partial_\phi \Phi_\pm)$ profiles are shown in Figs.~\ref{Fig_ExpPhiMorb}(c) and \ref{Fig_ExpPhiMorb}(d), which match reasonably with the simulated $M_{orb}^\pm$ profiles of Figs.~\ref{Fig_SimPhiMorb}(c) and \ref{Fig_SimPhiMorb}(d) respectively. The proportionality of $\mathcal{I}_\pm (\partial_\phi \Phi_\pm)$ to $M_{orb}^\pm$ is understood via Eq.~(\ref{Morb1}); and these results verify the correctness of our simulation and experiment.

Using $A_1 = 3.42239$ V/m and $A_2 = 1.44292$ V/m in Eq.~(\ref{lorbpm_Brew}), we obtain the results $l_{orb}^\pm = \mp 0.71596 \, \hbar$. Using the experimental $\mathcal{I}_\pm$ and $\Phi_\pm$ data in Eq.~(\ref{lorbPM_Exp}), we obtain the results $l_{orb}^{+ \, (Exp)} = -0.738 \, \hbar$ and $l_{orb}^{- \, (Exp)} = +0.740 \, \hbar$; which differ from the corresponding theoretical results by fractional errors $0.0312$ and $0.0340$ respectively. This is a reasonable match, considering the noise in the experimental data. Because of noise in the experimentally obtained $\Phi_\pm$ profiles, the numerically obtained derivatives $\partial_\phi \Phi_\pm$ are even more noisy, and lead to the above errors in the $l_{orb}^\pm$ measurements. These errors can be minimized by taking appropriate steps to reduce noise in the data, both experimentally and computationally. For the purpose of the present paper, the above results sufficiently verify our approach for the OAM per photon measurement.


\section{Conclusion} \label{Sec_Conc}

In the present paper, we have established a formalism for experimental determination of the OAM of a general 2D complex beam-field by using Barnett's AM flux density formalism. Our formalism involves phase measurements based on Stokes parameters in contrast with the standard ones based on interference patterns. We have demonstrated our OAM measurement by simulating an optical system, and also by creating it experimentally. 
The experimental results are in agreement with the simulated results, thus verifying the validity of our method. 

To our knowledge, our work is the first reporting of the measurement of OAM per photon of a beam-field implementing Barnett's formalism and Stokes parameter based phase measurements.
As compared to the widely used AM density formalism, the AM flux density formalism has met with quite limited applications in the recent literature. In this context, our present work also serves the purpose of exploring a significant application of the formalism in OAM measurements.
Our method will find applications in nano-optical processes and light-matter interaction involving the transfer of OAM to particles and across interfaces between different media.

\begin{acknowledgments}
A.D. thanks CSIR (India) for Senior Research Fellowship. N.K.V. thanks SERB (DST, India) for financial support.
\end{acknowledgments}


\bibliography{AD_NKV_BarnettStokesSPIE_Refs}

\end{document}